\documentclass[
  twocolumn,
  prb,
  showpacs,
  hyperref,
  amsmath,
  amssymb,
  superscriptaddress,
  nofootinbib,
  floatfix,
  longbibliography
]{revtex4-1}

\usepackage{mathtools}
\usepackage{bm}
\usepackage{dsfont}
\usepackage{graphicx}
\usepackage{pifont}
\usepackage{textcomp}
\usepackage{gensymb}
\usepackage{accents}
\usepackage{upgreek}
\usepackage{array}
\usepackage{hhline}
\usepackage{tabularx}
\usepackage{color}

\newcommand{\s}{\sum\limits}

\newcommand{\pa}{\partial}

\newcommand{\be}{\begin{equation}}
\newcommand{\e}{\end{equation}}
\newcommand{\beml}{\begin{subequations}}
\newcommand{\eml}{\end{subequations}}
\newcommand{\beq}{\begin{eqnarray}}
\newcommand{\eq}{\end{eqnarray}}
\newcommand{\ba}{\begin{array}}
\newcommand{\ea}{\end{array}}
\newcommand{\bpm}{\begin{pmatrix}}
\newcommand{\epm}{\end{pmatrix}}
\newcommand{\bc}{\begin{cases}}
\newcommand{\ec}{\end{cases}}
\newcommand{\lt}{\left}
\newcommand{\rt}{\right}
\newcommand{\n}{\nonumber}
\newcommand{\la}{\langle}
\newcommand{\ra}{\rangle}
\newcommand{\ep}{\varepsilon}

\newcommand{\bs}{\mathbf}
\newcommand{\bb}{\boldsymbol}

\DeclareMathOperator{\rot}{rot}
\DeclareMathOperator{\sign}{sign}

\begin{document}

\title{Orbital magnetization from interface reflections in a conductor with charge current}

\author{J.~Voss}
\affiliation{Radboud University, Institute for Molecules and Materials, NL-6525 AJ Nijmegen, The Netherlands}

\author{I.\,A.~Ado}
\affiliation{Radboud University, Institute for Molecules and Materials, NL-6525 AJ Nijmegen, The Netherlands}

\author{M.~Titov}
\affiliation{Radboud University, Institute for Molecules and Materials, NL-6525 AJ Nijmegen, The Netherlands}

\begin{abstract}
We propose that a high-quality flat interface or boundary can serve as a long-range skew scatterer for charged quasiparticles in a metal. When an electric current flows parallel to the interface, the balance between clockwise and counterclockwise reflections is disrupted, leading to a net orbital magnetization. This magnetization is maximized at the interface and varies linearly in the direction perpendicular to it. We suggest that this effect can be detected using spatially resolved Kerr effect measurements at distances up to the electron phase coherence length from the interface. Unlike the orbital Hall and orbital Edelstein effects, the proposed phenomenon does not require inversion symmetry breaking in the bulk of the sample and is unrelated to Hall effect physics.
\end{abstract}

\maketitle
Even though the spin Hall effect is widely regarded as a fundamental concept in spintronics \cite{Sinova2015review}, its relevance has been questioned by observations of spin Hall-like phenomena in materials with very weak or negligible spin-orbit coupling \cite{Yuasa2020}. To explain these observations, the concepts of the orbital Hall effect and the orbital Edelstein effect (also known as the orbital magnetoelectric effect) have been introduced \cite{Go_2021, atencia2024orbital}.  

The spin Hall and Rashba Edelstein effects, along with their orbital counterparts, can be classified as spectral effects. In other words, they require specific coupling terms in the effective Hamiltonians of Bloch electrons that break inversion symmetry -- these include spin-orbit couplings or couplings to the orbital momentum of Bloch electrons. Additionally, spin and orbital Hall effects rely on the assumption of transverse currents that are neither charge nor quasiparticle currents and do not satisfy conservation laws. The idea that such non-conserved currents could lead to an "accumulation" of magnetic moments at the boundaries is a hypothesis that has yet to be theoretically substantiated.  

\begin{figure}[tb]
\includegraphics[width=\columnwidth]{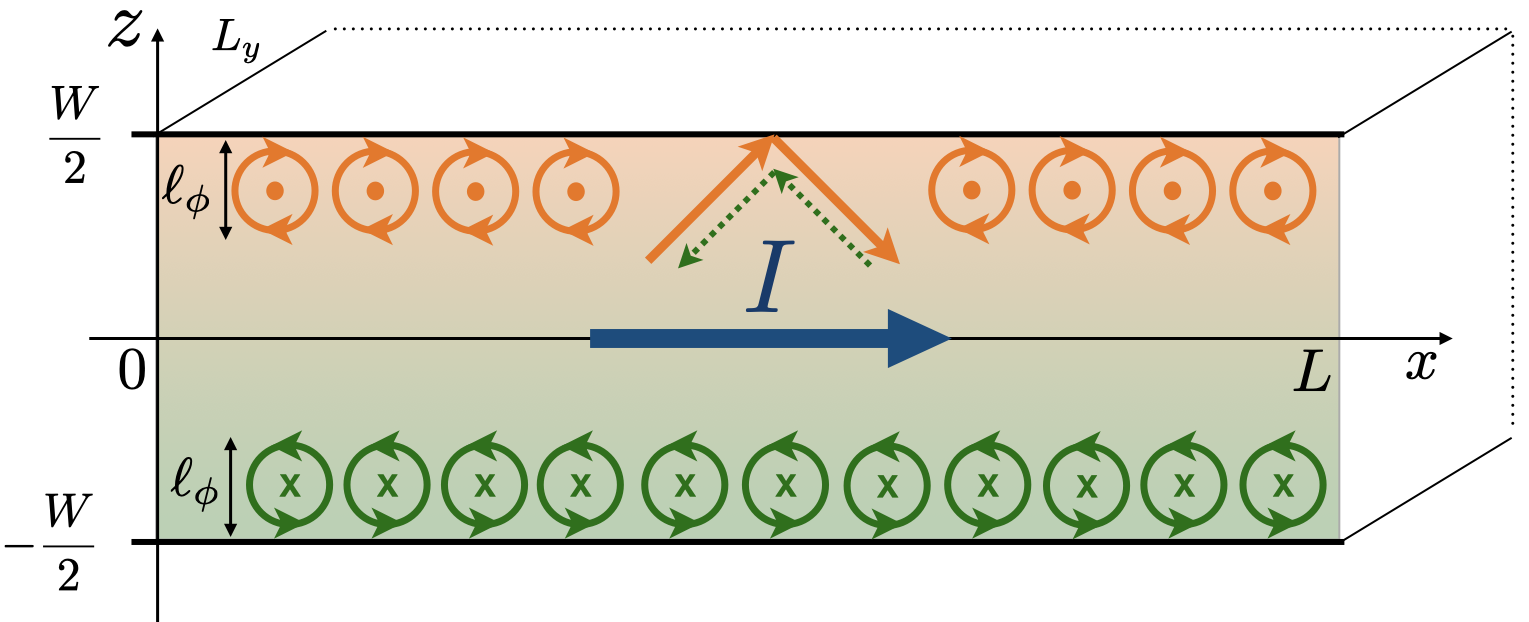}\\
\caption{Charge current flow in the $x$ direction induces a finite orbital moment density. Due to the charge current, extended states with $k_x > 0$ and $k_x < 0$ are populated asymmetrically, leading to an imbalance in the number of interface reflections where electron trajectories turn clockwise or counterclockwise around the $y$ direction. This imbalance results in an uncompensated orbital moment, which contributes to the non-diagonal optical susceptibility of the metal within a layer of thickness comparable to the dephasing length $\ell_\phi$ from the surface. In a symmetric sample, the effect has opposite signs at the top and bottom interfaces, ensuring that the total orbital magnetization of the sample remains zero.}
\label{fig:S}
\end{figure}

In this work, we adopt a different perspective and explore the possibility of orbital effects arising from quasiparticle scattering at a metal interface. A current flowing parallel to the interface corresponds to chaotic electron trajectories originating from the left reservoir and terminating in the right reservoir. Most of these trajectories inevitably undergo reflections from the metal boundaries. In the presence of charge flow, the populations of trajectories undergoing clockwise and counterclockwise reflections at a given interface become imbalanced, leading to a finite orbital magnetization in the vicinity of the interface. This orbital magnetization varies linearly with the coordinate perpendicular to the interface and can be detected through spatially resolved optical Kerr measurements \cite{Kato2004}. However, we argue that the symmetry-breaking effect induced by interface scattering cannot be observed in the optical Kerr response at distances exceeding the electron phase coherence length $\ell_\phi$ from the interface (see Fig.~\ref{fig:S}).  

Furthermore, we propose that the magnitude of the orbital edge magnetization is proportional to the product of the phase coherence length $\ell_\phi$ and the charge current density. This effect may provide an alternative explanation for the experimental observations of non-equilibrium edge magnetization in GaAs samples, which have traditionally been attributed to the spin Hall effect \cite{Kato2004}. From our perspective, the exceptionally long dephasing length of the electron gas in GaAs at $T=30$\,K, along with its high-quality boundaries defined by electrostatic gating, are key factors that have enabled the optical detection of non-equilibrium orbital magnetization over a large region near the sample edges (on the order of $14$\,$\mu$m). Notably, this value is consistent with estimates of the electron dephasing length in GaAs. Our estimation of the magnitude of this effect also qualitatively agrees with experimental observations.  

The scattering-induced mechanism of orbital magnetization in conductors is not a new concept. The idea that finite orbital magnetization can arise from the skew scattering of conduction electrons was first proposed in 1985 \cite{Levitov1985}. Indeed, one can argue that charge current flow in a metal containing special impurities, characterized by an asymmetric scattering cross-section, generates a finite orbital magnetization in the bulk. This effect is a direct consequence of the unequal probabilities of clockwise and counterclockwise rotations of electron trajectories upon impurity scattering and does not require any special coupling in the effective Hamiltonian of Bloch electrons. In our case, the entire metal interface can similarly be regarded as a macroscopic skew scatterer, inducing local orbital magnetization in the vicinity of the interface due to charge current flow.  

Circulating currents in diamagnetic conductors have been known since the early days of quantum mechanics \cite{Einstein1915}. Although these currents can be formally large, their total contribution to magnetization cancels out in thermodynamic equilibrium, except in cases such as persistent currents in nano-rings at ultralow temperatures \cite{persistent2009}. This cancellation arises due to the dense (continuous) spectrum of extended electron states in a metal. However, we argue that an electric current, by creating an imbalance in the population of left- and right-moving extended states, generates a finite local orbital magnetization through single-particle scattering states that reflect off the sample boundaries.  

By definition, extended electron states are the only states affected by a dc current. In linear response theory, one can therefore neglect the contribution of fully occupied bands and, more generally, all states far from the Fermi level. The expectation value of the operator $\bs{L}=\bs{r}\times\bs{v}$, where $\bs{r}$ is the position operator and $\bs{v}$ is the velocity operator, evaluated in an extended state yields the total angular momentum of the entire sample. However, this quantity is highly non-local, leading to well-known regularization issues in any conductor \cite{THONHAUSER}.  

In the simple conductor we consider below, the mean value of the operator $\bs{L}$ remains zero even in the presence of charge current (see Fig.~\ref{fig:S}). Nevertheless, a finite local orbital magnetization emerges. To describe this effect, one must introduce the thermodynamic density of orbital momentum, which does not have a direct relation to the operator $\bs{L}$.  

In thermodynamics, the total magnetic moment of a system is defined as the quantity conjugate to an external magnetic field $\bb{B}$. The differential of the grand potential $\Omega$ is given by $\mathrm{d} \Omega = - \bs{M} \cdot \mathrm{d} \bs{B}$, where $\bs{M}$ represents the total magnetic moment, and $\mathrm{d} \bs{B}$ is the differential of a homogeneous external magnetic field. A naive generalization of this relation would define magnetization as the functional derivative of the grand potential with respect to the local field $\bs{B}(\bs{r})$. However, such a definition is not entirely physical, as an elementary magnetic field exists only in the form of a flux line and cannot be represented as a three-dimensional delta function.  

Nevertheless, one can introduce the concept of orbital moment density per unit area by considering the variation of magnetic flux $\mathrm{d}\bb{\Phi}$ through a small two-dimensional (2D) surface (a flux tube). In this case, the conjugate variable corresponds to the square orbital moment density (i.e., the magnetic moment per unit 2D area in the plane perpendicular to the flux), leading to the relation $\mathrm{d} \Omega = - \bb{\mathcal{M}} \cdot \mathrm{d} \bb{\Phi}$. We use this definition to characterize the square density $\bb{\mathcal{M}}$ of orbital magnetic moments in a conductor.  

For definiteness, we consider a specific sample geometry illustrated in Fig.~\ref{fig:S}. The conducting layer has a thickness $W$ in the $z$ direction, a large (formally infinite) length $L_y$ in the $y$ direction, and a finite length $L$ in the $x$ direction. A charge current is applied along the $x$ direction. In this setup, a non-equilibrium orbital magnetization emerges, oriented along the $y$ direction.  

To define the orbital moment density, we consider an external magnetic field $\bs{B}$ in the form of a magnetic flux line:
\be
\bs{B}=\Phi_0\hat{\bs{y}}\,\delta(x-x_0)\delta(z-z_0),
\e
where $\hat{\bs{y}}$ is the unit vector in the $y$ direction, and $\Phi_0$ represents the magnetic flux penetrating any $(x, z)$ cross-section that includes the point $(x_0, z_0)$. The corresponding vector potential takes the form:
\be
\label{flux_tube}
\bs{A}(\bs{r})=\Phi_0\lt(\frac{1}{2\pi} \frac{\hat{\bs{y}}\times(\bs{r}-\bs{r}_0)}{|\hat{\bs{y}}\times(\bs{r}-\bs{r}_0)|^2} +\bb{\nabla}\chi\rt),
\e  
where $\chi$ is an arbitrary gauge function.

The vector potential enters the effective electron Hamiltonian via the Peierls substitution $\bs{p} \to \bs{p} - e\bs{A}$, where $e=-|e|$ is the electron charge, and we use natural units with $c=1$. The contribution to the electron free energy in the linear order with respect to the vector potential is given by:
\be
\delta\Omega =- \int d^3\bs{r}\;  \bb{j}(\bs{r})\cdot \bs{A}(\bs{r}),
\e
where $\bs{j}(\bs{r})$ denotes the charge current density. Differentiating $\delta\Omega$ with respect to $\Phi_0$ (evaluated at $\Phi_0=0$) yields the orbital moment density projected onto the $\hat{\bs{y}}$ direction:
\be
\label{Mdef}
\mathcal{M}(\bs{r}_0) =  -\int d^3\bs{r}\; \bs{j}(\bs{r})\cdot \lt(\frac{1}{2\pi} \frac{\hat{\bs{y}}\times(\bs{r}-\bs{r}_0)}{|\hat{\bs{y}}\times(\bs{r}-\bs{r}_0)|^2} +\bb{\nabla}\chi\rt),
\e 
where $\bs{r}_0=(x_0, 0, z_0)$ specifies the axis. The definition in Eq.~(\ref{Mdef}), derived from the variation of the local magnetic field, differs from the approach suggested in Ref.~\onlinecite{Resta2013}, which defines local orbital magnetization based on the density of the orbital moment operator $\bs{L}$. 

\begin{figure}[t]
\includegraphics[width=\columnwidth]{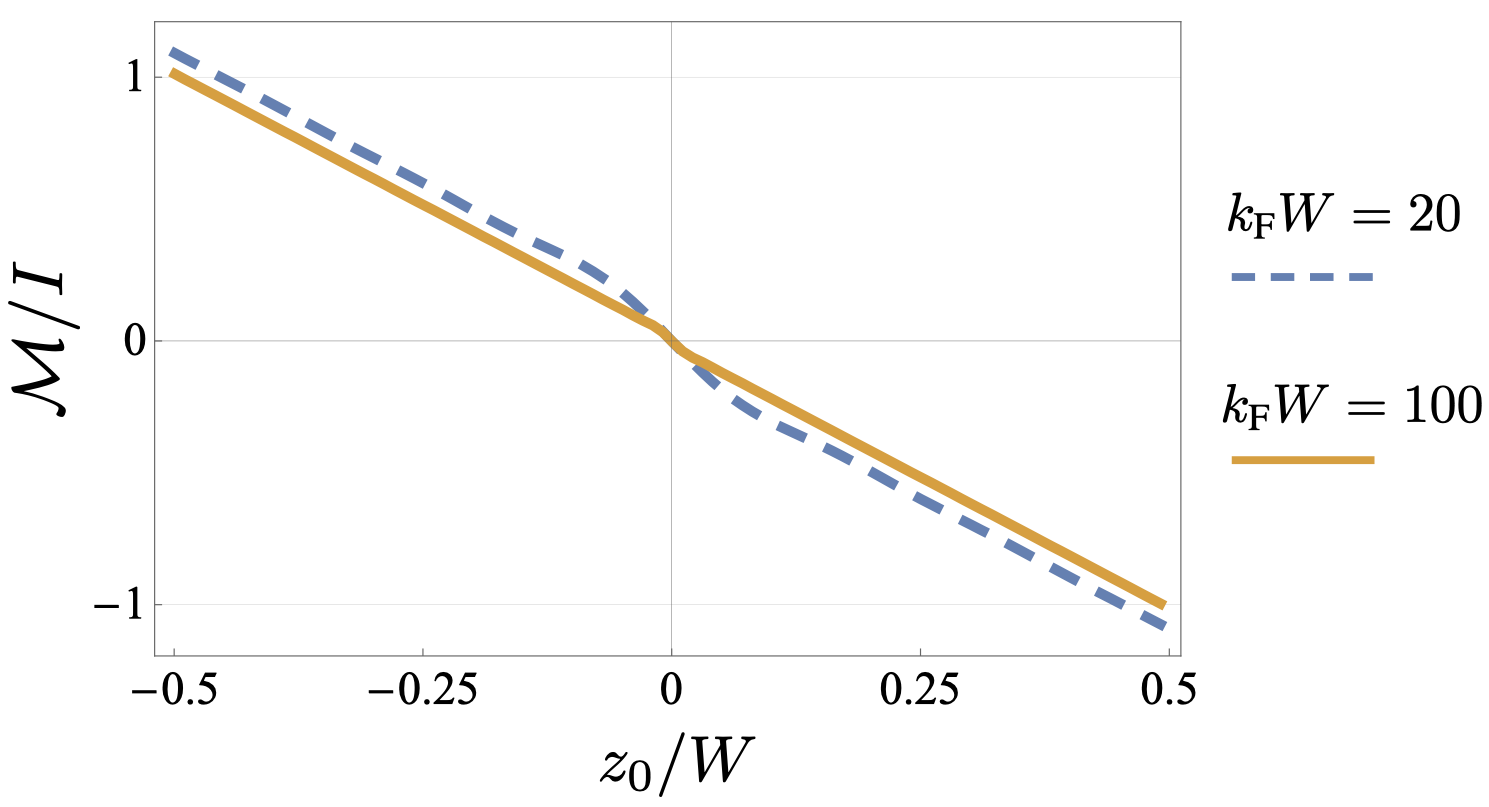}\\
\caption{The 2D orbital moment density profile of a ballistic wire, obtained from Eq.~(\ref{UU1}). The density reaches a value of $I/2$ at the top and bottom interfaces. The effect of Friedel oscillations vanishes in the limit $k_\textrm{F} W \gg 1$.} 
\label{fig:U}
\end{figure}

The total magnetic moment in the $\hat{\bs{y}}$ direction can be obtained by integrating the orbital moment density over a two-dimensional surface:
\be
\label{totalM}
M=\int dx_0 dz_0\, \mathcal{M}(x_0, z_0).
\e
This integration yields the standard result:
\be
\label{Mtotal}
M=  \frac{1}{2} \int d^3\bs{r}\; \hat{\bs{y}}\cdot \lt[\bs{r}\times\bs{j}(\bs{r})\rt] = \frac{e}{2} \int d^3\bs{r}\; \hat{\bs{y}}\cdot \lt\la \bs{L} \rt\ra,
\e
where $\lt\la \bs{L} \rt\ra$ is the expectation value of the orbital moment operator.

The total orbital moment in Eq.~(\ref{Mtotal}) does not capture the effect of edge orbital magnetization illustrated in Fig.~\ref{fig:S}. Moreover, symmetry considerations require that the total moment must vanish identically ($M=0$). The definition in Eq.~(\ref{Mtotal}) is also challenging to apply in conductors due to the unbounded nature of the position operator $\bs{r}$ \cite{Hirst_1997, Resta_2010, Thonhauser_1995, Niu_2005, Niu_2007, THONHAUSER}. In contrast, the orbital moment density of Eq.~(\ref{Mdef}) is well-defined and does not require additional regularization. While Eq.~(\ref{totalM}) must hold by construction, it may be violated in practise if the regularization of Eq.~(\ref{totalM}) does not respect gauge symmetry.

To illustrate the application of Eq.~(\ref{Mdef}), we evaluate the local orbital magnetization in a quantum wire with charge flow. We employ an effective model that does not resolve individual atomic orbitals but instead operates with electron envelope wave functions that vary smoothly on atomic scales. The simplest effective Hamiltonian of this kind is the Schrödinger equation $H\Psi = E\Psi$, where:
\be
H = \frac{(p - e\bs{A})^2}{2m} + V(\bs{r}),
\e
with $e=-|e|$ as the electron charge, $c=1$, and $V(\bs{r})$ representing an external or disorder potential. We assume the geometry of Fig.~\ref{fig:S} with hard-wall boundary conditions $\Psi(z=\pm W/2) = 0$. 

Using the Landauer-Büttiker formalism, we construct the density matrix for an open system with charge current flow. In a ballistic system (i.\,e., with no disorder, $V=0$), the density matrix can be expressed in terms of the scattering states originating from reservoirs with different chemical potentials. For the sample geometry in Fig.~\ref{fig:S}, we define the left- and right-moving scattering states as:
\beml
\begin{align}
&\Psi_{n,s,>}=\sqrt{\frac{2}{W L_y v_x}}e^{ikx}e^{ip_s y}\sin(q_n (z+W/2)),\\
&\Psi_{n,s,<}=\sqrt{\frac{2}{W L_y v_x}}e^{-ikx}e^{ip_sy}\sin(q_n (z+W/2)),
\end{align}
\eml
where $q_n = \pi n/W$ for $n=1, 2, 3, \dots$. The quantization follows from the hard-wall boundary conditions in the $z$ direction. In this model, $v_x = \hbar k/m$ and $k = \sqrt{2mE/\hbar^2 - q_n^2 - p_s^2} > 0$. Assuming $L_y \gg W$, the boundary conditions in the $y$ direction become irrelevant. Here, we impose periodic boundary conditions $\Psi(y=0) = \Psi(y=L_y)$, leading to quantized momenta $p_s = 2\pi s/L_y$ with arbitrary integer values of $s$.

The scattering states are normalized to a unit charge current flux. The current density for a given state $\Psi$ is:
\be
\label{jjj}
J_x[\Psi_>] = -J_x[\Psi_<] = \frac{2 e}{L_y W} \sin^2(q_n (z+W/2)).
\e
The total charge flux of a scattering state is given by: \cite{Note1}
\be
Q[\Psi]=\int_{-W/2}^{W/2}\textrm{d}z\,\int_{0}^{L_y}\textrm{d}y\,J_x[\Psi],
\e
which confirms the normalization condition $Q[\Psi_>] = e$ and $Q[\Psi_<] = -e$.

In the Landauer-Büttiker picture, scattering states are populated according to the Fermi-Dirac distributions of the left and right reservoirs. The total current flowing from left to right is:
\begin{align}
I &=\s_{s,n} \int\frac{\textrm{d}E}{2\pi\hbar}\, \lt(Q[\Psi_{n,s,>}]f_L(E)+Q[\Psi_{n,s,<}]f_R(E)\rt)\n\\
&=e\int\frac{\textrm{d}E}{2\pi\hbar} (f_L(E)-f_R(E)) N(E),
\label{CQ}
\end{align}
where $N(E)$ represents the number of available conduction channels:
\be
N(E) = \frac{W L_y}{\pi} k_\textrm{F}^2.
\e
Equation~(\ref{CQ}) is nothing but the well-known Landauer formula for ballistic waveguides.

In the linear response regime, one can use
\be
\label{red}
f_L(E)-f_R(E)=f(E_\textrm{F}-eV_\textrm{bias})-f(E_\textrm{F})\approx eV_\textrm{bias} \lt(- \frac{\pa f}{\pa E}\rt),\n
\e
where $E_\textrm{F}$ is the Fermi energy, and $V_\textrm{bias}$ is the applied voltage bias. For practical purposes, we can approximate $- \pa f/\pa E$ by a delta function, $\delta(E-E_\textrm{F})$, yielding:
\be
\label{tot}
I= \frac{e^2}{4\pi h} W L_y k_\textrm{F}^2 V_\textrm{bias}=\frac{2e^2}{h} W L_y \nu E_\textrm{F} V_\textrm{bias},
\e
where $\nu=m/\pi\hbar^2$ is the two-dimensional density of states. The dimensionless quantity $W L_y \nu E_\textrm{F} \gg 1$ quantifies the sample conductance.

Before evaluating the local orbital moment in the wire, we note that, due to symmetry considerations, it depends only on the transverse coordinate $z$. To define the corresponding one-dimensional density, we consider a magnetic field of the form $\bs{B}=\hat{\bs{y}}(\Phi_0/L) \delta(z-z_0)$. The corresponding vector potential, in the Landau gauge, can be chosen as $\bb{A}=\hat{\bs{x}} (\Phi_0/2L) \sign(z-z_0)$. The orbital magnetic moment square density is then given by:
\begin{align}
\mathcal{M}=& \sum_{s,n}\int d^3\bs{r} \int\frac{\textrm{d}E}{2\pi\hbar}\; \frac{1}{2L} \sign(z-z_0) \n\\
&\times \lt(f_L(E) J_x[\Psi_{n,s,>}]  + f_R(E) J_x[\Psi_{n,s,<}]\rt),
\label{MQ}
\end{align}
where the summation runs over all scattering states.

Using the result from Eq.~(\ref{jjj}), we obtain:
\begin{align}
\mathcal{M}=&\frac{e}{L L_y W}  \sum_{s,n}\int d^3\bs{r} \int \frac{dE}{2\pi \hbar}\; \lt(f_L(E)-f_R(E)\rt)\n\\
&\times \sign(z-z_0) \sin^2\lt(q_n (z+W/2)\rt).
\label{cool}
\end{align}

To relate the density $\mathcal{M}$ to the total current, we use Eq.(\ref{tot}) and project the scattering states onto the Fermi level using Eq.(\ref{red}). The summation over the channels in Eq.(\ref{cool}) is evaluated as:
\begin{align}
&2\sum_{s=-\infty}^\infty \sum_{n=1}^\infty  \sin^2\lt(q_n (z+W/2)\rt)\, \Theta(q_n^2+p_s^2 \leq 2mE/\hbar^2)\n\\
&\qquad =N(E)\,\lt(1+\frac{J_1(2k_\textrm{F}z)}{k_\textrm{F}z}\rt),
\end{align}
where $J_1(x)$ is the Bessel function, and $k_\textrm{F}=\sqrt{2mE_\textrm{F}/\hbar^2}$ is the Fermi momentum. By integrating over $x$ and $y$ in Eq.(\ref{cool}), we obtain:
\be
\mathcal{M}= \frac{I}{2W} \int_{-W/2}^{W/2} dz\;\lt(1+\frac{J_1(2k_\textrm{F}z)}{k_\textrm{F}z}\rt)\sign(z-z_0),
\label{UU1}
\e
which is illustrated in Fig.~\ref{fig:U}.

Thus, we find that the total magnetic moment $M$ vanishes due to the exact top-bottom symmetry $z \to -z$, while the orbital moment density $\mathcal{M}$ remains finite. The term $J_1(2k_\textrm{F}z)/k_\textrm{F}z$ accounts for Friedel oscillations originating from the sample boundaries. These oscillations cause small deviations of $\mathcal{M}$ from a linear dependence on $z_0$. In the limit $W k_\textrm{F}\gg 1$, the effect of Friedel oscillations becomes negligible, and the integration over $z$ simplifies to yield $\mathcal{M}= -I z_0/W$.

The usual (three-dimensional) magnetization of the system is defined as $\bs{\mathcal{M}}^\textrm{3D}=\bs{\mathcal{M}}/L_y$. Therefore, our the local orbital magnetization in the ballistic wire is simply given by $\bb{\mathcal{M}}^\textrm{3D}(z) = -\overline{j} z \hat{\bs{y}}$, where $\overline{\bs{j}} =\hat{\bs{x}} I /L_yW$ is the mean current density. This result clearly satisfies the Maxwell relation $\overline{\bs{j}}= \rot \bb{\mathcal{M}}^\textrm{3D}$. Hence, in a ballistic setup, the entire charge current must be regarded as a magnetization current rather than a current of free charges. One may think of the ballistic setup as a large artificial atom. When right-going single-particle states are populated more than left-going ones, a local orbital moment density (which is linear in the transversal direction) emerges due to symmetry breaking at the boundaries.

\begin{figure}[tb]
\includegraphics[width=0.95\columnwidth]{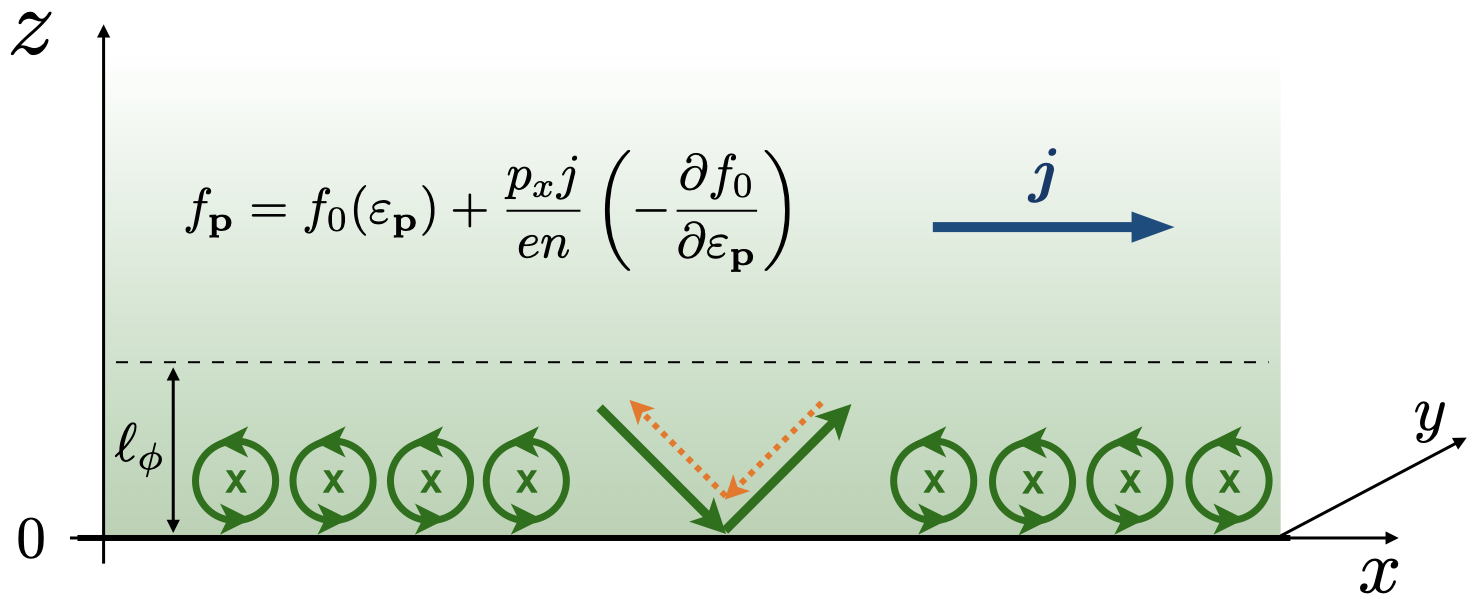}\\
\caption{Formation of orbital edge magnetization in a conductor with a current flow.}
\label{fig:G}
\end{figure}

In the case of a disordered system, both Eqs.~(\ref{CQ}) and (\ref{MQ}) are expressed through the reflection and transmission amplitudes. However, the result in Eq.~(\ref{UU1}), which essentially reduces to $\bb{\mathcal{M}}^\textrm{3D}(z) = -\overline{j} z \hat{\bs{y}}$, remains valid. It is believed that this non-equilibrium orbital magnetization can be probed locally using optical Kerr rotation. The latter is linked to the local Hall conductivity $\sigma_{xy}(\omega)$ at an optical frequency $\omega$, measured in the presence of a dc current. The dc current modifies the electron distribution function in the bulk of the sample as
\be
f(\ep_{\bs{k}},\hat{\bs{k}})=f_0(\ep_{\bs{k}})+\frac{j \hbar k_x}{e n}\lt(-\frac{\pa f_0(\ep_{\bs{k}})}{\pa \ep_{\bs{k}}}\rt),
\e
where $n$ is the charge density and $j$ is the current density in the $x$ direction. In other words, the current flow populates more states with positive $k_x$ than those with negative $k_x$ (see Fig.~\ref{fig:G}).

In a system without boundaries, the optical response $\sigma_{xy}(\omega)$ is zero by symmetry, as the only distinguished direction in this system is $x$. Nevertheless, it has been shown in Ref.~\onlinecite{Levitov1985} that in the presence of skew scattering, a finite response (proportional to $j$) can appear. In our system, the sole origin of skew scattering is boundary reflection. For a local Kerr measurement to be finite, the phase shift induced by boundary scattering in electron trajectories originating from and returning to the measurement area must be detected. This is analogous to the working principle of a sonar, which measures the distance from a moving ship to the seabed using phase shifts in sound waves. Interactions randomize the phase of single-particle electron states on distances longer than the dephasing length $\ell_\phi$. Thus, the Kerr rotation angle must in any case vanish at distances larger than $\ell_\phi$ from the metal interface. In the experiments of Kato \emph{et al.} \cite{Kato2004}, a clear linear dependence of Kerr rotation on $z$ was observed, extending up to approximately 14\,$\rm{\mu}$m from the sample edge. The latter scale does indeed coincide with the electron dephasing length in these samples.

A microscopic theory of local Kerr measurements in a sample with charge flow is rather complex and will be presented elsewhere. Here, we rely on a qualitative argument. We define a surface layer of the system as the region within a width $\ell_\phi$ from the interface, where electrons retain phase coherence and a single-electron picture remains valid.

The surface layer can be modeled as a phase-coherent sample of width $\ell_\phi$, with the condition $\mathcal{M}=0$ at a depth $\ell_\phi$ from the interface. In the natural limit $k_\textrm{F} \ell_\phi \gg 1$, Friedel oscillations can be ignored, while the orbital moment density $\mathcal{M}$ can be assumed to increase linearly from $0$ at a distance $\ell_\phi$ from the surface to $\pm I \ell_\phi/W$ at the surface (where $I \ell_\phi/W$ represents the fraction of the total current flowing through the surface layer).

Consequently, the total orbital magnetic moment of the surface layer is given by $M_\textrm{edge}=\pm I \ell_\phi^2 L/2W$. This result suggests that the characteristic electron orbit area contributing to the edge effect is $\ell_\phi^2$. This characteristic orbital area is significantly larger than the unit cell, posing a challenge for any \emph{ab initio} (atomistic) approach to the phenomenon \cite{THONHAUSER}.

\begin{figure}[tb]
\includegraphics[width=0.8\columnwidth]{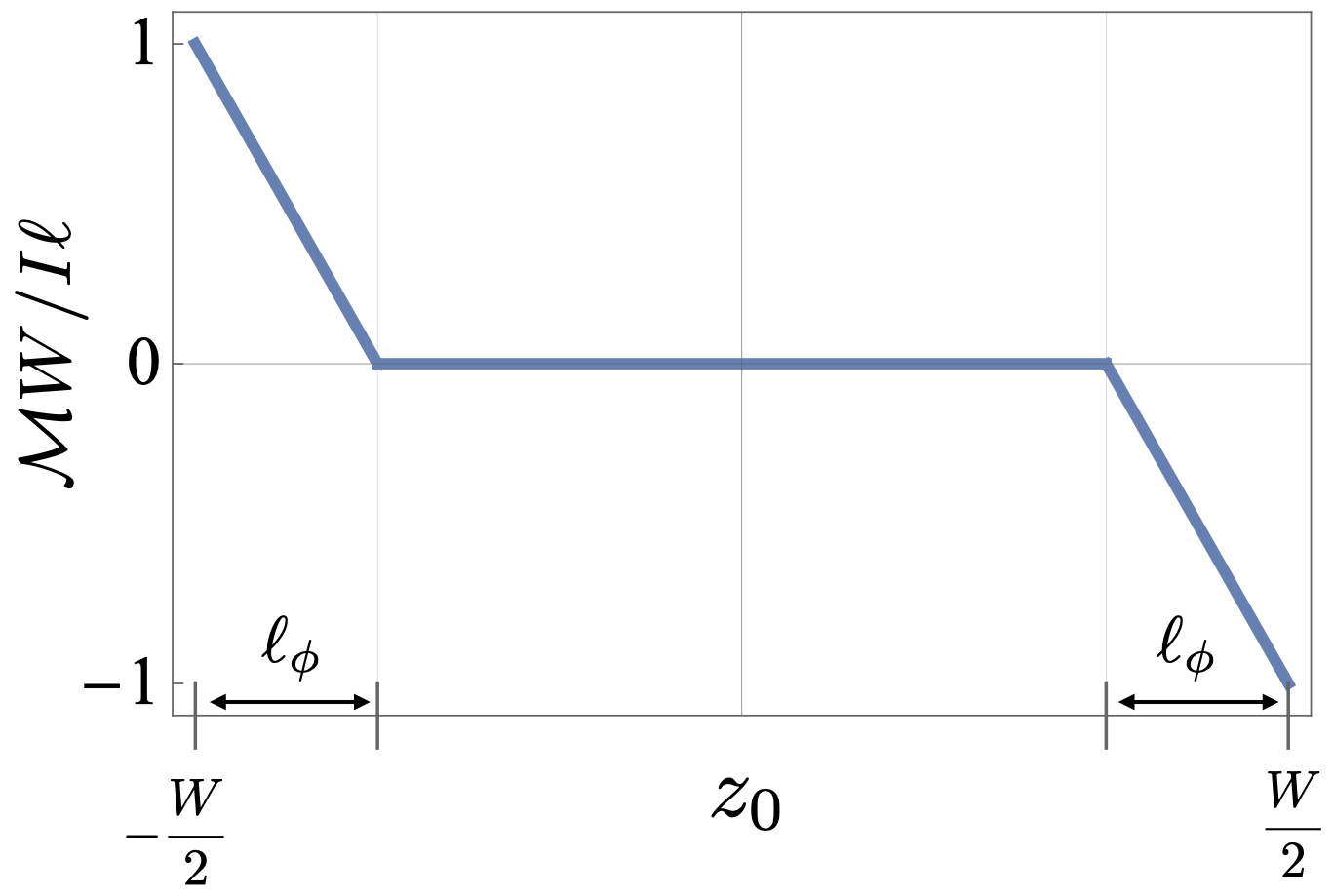}\\
\caption{The sketch of orbital moment density $\mathcal{M}$ profile measured by Kerr rotaiton.}
\label{fig:V}
\end{figure}

The resulting orbital moment density is illustrated in Fig.~\ref{fig:V}. It is useful to introduce the average current density $\bar{j}$ using $I=L_yW\, \bar{j}$, yielding $M_\textrm{edge}=\bar{j}\ell_\phi^2LL_y/2$. In terms of standard three-dimensional magnetization, we obtain
\be
\label{est}
\mathcal{M}^\textrm{3D}_\textrm{edge}=\pm\frac{M_\textrm{edge}}{2L_yL\ell_\phi}=\pm\bar{j}\ell_\phi/2,
\e
which is a remarkably large effect that does not involve spin-orbit coupling. By contrast, the Rashba-Edelstein and spin Hall effects \cite{Rashba_Edelstein, Hirsch99, Sinova2015review}, which share the same symmetry, necessarily include additional small factors $\Delta_\textrm{so}/E_\textrm{F}$ and $(\Delta_\textrm{so}/E_\textrm{F})^2$, respectively, where $\Delta_\textrm{so}$ is the spin-orbit splitting. These factors range from $10^{-4}$ to $10^{-9}$, significantly diminishing the impact of these effects.

In the experiment of Ref.~\onlinecite{Kato2004}, the charge current density reached $\bar{j}=50\, \rm{\mu}$A$/\rm{\mu}$m$^2$. Thus, for the dephasing length $\ell_\phi \simeq 14$\,$\rm{\mu}$m, the effective edge magnetization can be estimated as $\mathcal{M}^\textrm{3D} \simeq 700$\,A$/$m $\approx 8.8$\,Oe. This value is only three orders of magnitude smaller than the saturation magnetization of yttrium iron garnet (YIG).

In Ref.~\onlinecite{Kato2004}, the edge of the electron gas was defined via electrostatic gating, introducing a smooth confinement potential extending over $300$\,nm. This reduced the effect magnitude by an order of magnitude, yet the total orbital magnetic moment at the boundary agreed well with our estimate. Thus, the result in Eq.~(\ref{est}) is consistent with the observed edge magnetization in GaAs samples carrying charge current \cite{Kato2004}.

The discussion of the type of magnetization detected by Kerr rotation in a metal with current flow is closely related to the somewhat artificial problem of decomposition of total charge current density in a conductor into the net current density of free charges $\bs{j}_f$ and magnetization current, which is typically associated with localized electrons, $\overline{\bs{j}}= \bs{j}_f+ \rot \bb{\mathcal{M}}^\textrm{3D}$. We propose that this decomposition may have a meaningful physical interpretation in the context of magnetization measured through optical methods. Specifically, the magnetization current can be understood as the single-particle (i.\,e. phase-coherent) contribution, arising from electron trajectories shorter than the dephasing length, whereas the free charge current constitutes a non-coherent component devoid of quantum phase information. Reflection at the interface, combined with a charge current parallel to it, creates an imbalance in the population of left- and right-moving single-particle scattering states within the surface layer. This imbalance is perceived by Kerr rotation as local magnetization.  

The mechanism discussed above may offer an alternative explanation for experimental results that are often attributed to the direct manifestation of the spin Hall effect~\cite{Kato2004}. Likewise, non-equilibrium orbital edge magnetization from skew-scattering off the boundary could account for observations currently ascribed to the orbital Hall effect~\cite{OHE_nature, OHE_Kawakami, Hanle23, ANDO23, choi23}.  

In conclusion, we have analyzed the emergence of an orbital magnetic moment at the surface of a conducting sample in the presence of a parallel charge current. This effect arises due to an imbalance in surface reflections between quasiparticles with $k_x > 0$ and $k_x < 0$ and does not require spin-orbit coupling. The phenomenon shares the same symmetry as the spin Hall and Rashba-Edelstein effects but is orders of magnitude stronger. The resulting non-equilibrium edge magnetization can reach values of $\bar{j} \ell_\phi/2$, where $\ell_\phi$ is the electron dephasing length, and $\bar{j} = I/L_y W$ represents the charge current density in the sample. This effect is sensitive to interface quality and is suppressed in the presence of interface disorder. It provides an alternative explanation for some experimental observations currently attributed to the orbital Hall, Rashba-Edelstein, and spin Hall effects. A microscopic theory of this effect can be developed using the approach outlined in Refs.~\onlinecite{Levitov1985, Ado2024}.

We appreciate discussions with Vladimir Bashmakov, Dimi Culcer, Rembert Duine, Ivan Iorsh, Mikhail Katsnelson, Yuriy Mokrousov and Thierry Valet. 

We acknowledge funding from the European Union’s Horizon 2020 research and innovation programme under the Marie Sklodowska-Curie grant agreement No 873028.

\bibliography{biblio}

\end{document}